\documentclass[%
 reprint,
 amsmath,amssymb,
 aps,
]{revtex4-2}

\usepackage{graphicx}
\usepackage{dcolumn}
\usepackage{bm}

\usepackage[colorlinks=true,linkcolor=blue,anchorcolor=blue,citecolor=blue, filecolor=blue,urlcolor=blue,bookmarksnumbered=true,pdfview=FitB]{hyperref}

\usepackage{subfig}
\usepackage{mathrsfs}
\usepackage{float}
\usepackage{url}
\graphicspath{{figures}}

\allowdisplaybreaks

\begin{document}

\preprint{APS/123-QED}

\title{Transverse force tomography inside a proton from Basis Light-front Quantization}

\author{Ziqi Zhang}
\email{zhangziqi@impcas.ac.cn}
\affiliation{Institute of Modern Physics, Chinese Academy of Sciences, Lanzhou 730000, China}
\affiliation{School of Nuclear Science and Technology, University of Chinese Academy of Sciences, Beijing 100049, China}
\affiliation{CAS Key Laboratory of High Precision Nuclear Spectroscopy, Institute of Modern Physics, Chinese Academy of Sciences, Lanzhou 730000, China}

\author{Chandan Mondal}
\email{mondal@impcas.ac.cn}
\affiliation{Institute of Modern Physics, Chinese Academy of Sciences, Lanzhou 730000, China}
\affiliation{School of Nuclear Science and Technology, University of Chinese Academy of Sciences, Beijing 100049, China}
\affiliation{CAS Key Laboratory of High Precision Nuclear Spectroscopy, Institute of Modern Physics, Chinese Academy of Sciences, Lanzhou 730000, China}

\author{Siqi Xu}
\email{xsq234@impcas.ac.cn}
\affiliation{Institute of Modern Physics, Chinese Academy of Sciences, Lanzhou 730000, China}
\affiliation{School of Nuclear Science and Technology, University of Chinese Academy of Sciences, Beijing 100049, China}
\affiliation{CAS Key Laboratory of High Precision Nuclear Spectroscopy, Institute of Modern Physics, Chinese Academy of Sciences, Lanzhou 730000, China}

\author{Xingbo Zhao}
\email{xbzhao@impcas.ac.cn}
\affiliation{Institute of Modern Physics, Chinese Academy of Sciences, Lanzhou 730000, China}
\affiliation{School of Nuclear Science and Technology, University of Chinese Academy of Sciences, Beijing 100049, China}
\affiliation{CAS Key Laboratory of High Precision Nuclear Spectroscopy, Institute of Modern Physics, Chinese Academy of Sciences, Lanzhou 730000, China}

\author{James P. Vary}
\email{jvary@iastate.edu}
\affiliation{Department of Physics and Astronomy, Iowa State University, Ames, IA 50011, U.S.A.}

\collaboration{BLFQ Collaboration}

\date{\today}

\begin{abstract}

The twist-3 transverse spin--dependent nucleon structure function $g_2$ arises in 
high-energy processes involving a transversely polarized nucleon. Its connection 
to quark--gluon correlations allows for an interpretation in terms of the average 
transverse color Lorentz force acting on unpolarized quarks inside a transversely 
polarized nucleon. In this work, we investigate this force using light-front wave 
functions obtained by diagonalizing the light-front Hamiltonian with quantum 
chromodynamics inputs within the Basis Light-front Quantization approach. We evolve 
our results to a common scale of $5~\mathrm{GeV}^2$ and present the corresponding 
form factors in momentum space as well as the transverse force components in 
impact-parameter space. These distributions provide a complementary perspective on 
the Sivers asymmetry in transversely polarized deep-inelastic scattering. 
In the forward limit, we extract the twist-3 reduced matrix  element $d_2$, 
and our results are found to be comparable with those from other 
theoretical calculations and experimental determinations.
\end{abstract}

\maketitle


\section{\label{sec:1} Introduction}
In modern nuclear and particle physics, high-energy scattering processes serve as a cornerstone for unraveling the structure of hadrons, testing quantum chromodynamics (QCD), and exploring the fundamental interactions of matter. As an important process among them, deep inelastic scattering (DIS) probes the internal structure of nucleons and nuclei, enabling the extraction of parton distribution functions (PDFs)—a set of fundamental quantities that quantify the probability of finding a parton with a given fraction of the nucleon’s momentum. To analyze the invariant matrix element of the scattering process, the operator product expansion (OPE) and twist analysis are employed \cite{jaffe2007spin}, where $twist$ is defined as the dimension of the operator minus the dimension of the spin.

The parton model captures only leading-order (twist-2) contributions and neglects higher-twist effects. In certain cases, however, partonic intuition can be extended to higher twist, such as relating twist-three operators to quark--gluon correlations~\cite{Shuryak:1981pi}, which admit a phenomenologically appealing interpretation in terms of a color--Lorentz force~\cite{aslan2019transverse,PhysRevD.88.114502,Crawford:2024wzx,Liu:2025ypg}. In this work, we compute the transverse impact-parameter--space distribution of this color--Lorentz force directly from light-front wave functions (LFWFs) within the Basic Light-front Quantization (BLFQ) framework~\cite{Vary:2009gt,Xu:2023nqv}. This requires extending the twist-three forward matrix element $d_2$ to off-forward kinematics and introducing three form factors (FFs) that encode the force distributions~\cite{aslan2019transverse,Crawford:2024wzx,Liu:2025ypg}. We show that these distributions provide a consistent and intuitive framework for understanding single-spin asymmetries in semi-inclusive deep-inelastic scattering (SIDIS).

The matrix element associated with the axial-vector current is of particular interest, which is defined as \cite{jaffe1992chiral}
\begin{align}
    \int \frac{d\lambda}{2\pi} e^{i\lambda x} \langle PS|&\bar{\psi}(0) \gamma^\mu \gamma_5 \psi(\lambda n) |PS \rangle = 2[g_1(x)p^\mu(S\cdot n) \notag \\
    &+ g_T(x) S^\mu_\perp + M^2 g_3(x)(S\cdot n)n^\mu],
\end{align}
where \( p^\mu \) and \( n^\mu \) are light-like vectors along the `\( - \)' and `\( + \)' light-cone directions with \( p\cdot n = 1 \), and \( S^\mu \) denotes the proton polarization vector. These structure functions contribute to hard processes at the orders of \( g_1(x) \), \( g_T(x)/Q \), and \( g_3(x)/Q^2 \), and thus correspond to twist-2, twist-3, and twist-4, respectively. Here, \( x \) is the Bjorken scaling variable, and \( Q^2 \) is the momentum transfer from the virtual photon.

The presence of \( S^\mu_\perp \) implies that the twist-3 structure function \( g_T(x) \) can be cleanly accessed in transversely polarized protons. This is particularly significant because \( g_T(x) \) is sensitive to quark--gluon correlations~\cite{jaffe1991studies,shuryak1982theory}, which are in turn connected to transverse spin asymmetries and the transverse color Lorentz force~\cite{aslan2019transverse,PhysRevD.88.114502}. However, it should be noted that \( g_T(x) \) does not admit a simple single-particle density interpretation.

Experimental data exist for the chirally even spin-dependent twist-3 parton distribution $g_2(x)=g^q_T(x) - g^q_1(x)$, including results from polarized DIS experiments at SLAC (E142 \cite{anthony1996deep}, E143 \cite{abe1996measurements,abe1998measurements}, E154 \cite{abe1997measurement}, E155 \cite{arnold1999measurement,anthony2003precision}), HERMES \cite{hermes2012measurement}, JLab (JAM \cite{sato2016iterative}, Hall A \cite{zheng2004precision,flay2016measurements}, RSS \cite{wesselmann2007proton}), and CERN (SMC \cite{ballintijn1997spin}) (alongside virtual photon asymmetries). Additionally, the reduced twist-3 matrix element $d_2$ has been measured at HERMES \cite{hermes2012measurement}, JLab \cite{flay2016measurements}, SANE \cite{armstrong2019revealing}.

On the theoretical side, the twist-3 matrix element $d_2$ has been computed directly in lattice QCD~\cite{burger2022lattice,crawford2025transverse,liu2025color}. Significant attention has also been devoted to the transverse color Lorentz force, which provides an intuitive physical interpretation of twist-3 effects in terms of quark--gluon interactions. In particular, the QCDSF collaboration~\cite{crawford2025transverse} and studies based on the instanton liquid model~\cite{liu2025color} have computed the spatial distribution of this force in impact-parameter space. These developments are highly relevant for future precision studies at the planned Electron--Ion Collider (EIC) in the United States~\cite{khalek2022science} and the proposed Electron--Ion Collider in China (EicC)~\cite{anderle2021electron}.


The structure of this work is as follows. In Sec. \ref{sec:2}, we briefly introduce the BLFQ framework and present the parameters employed in this calculation. In Sec. \ref{sec:3}, we elaborate on the definition of the transverse color Lorentz force, its connection to the twist-3 spin-dependent structure function $g_2$, as well as their overlap representations adopted for the numerical calculations. In Sec. \ref{sec:4}, we present our numerical results, and discussions are provided in Sec. \ref{sec:5}.

\section{\label{sec:2} Basis light front quantization}
BLFQ is a non-perturbative method for solving the eigenvalue problem of light-front (LF) QCD, which incorporates LF QCD interactions via an effective LF Hamiltonian. The core eigenvalue equation in BLFQ is
\begin{equation}
P^- P^+ \, |\psi\rangle = M^2 \, |\psi\rangle ,
\end{equation}
where $P^\pm = P^0 \pm P^3$. Here, $P^+$ denotes the total longitudinal momentum of the system, $P^-$ is the light-front Hamiltonian, $M^2$ is the invariant mass squared of the eigenstate, and $|\psi\rangle$ is the corresponding light-front eigenstate.
For the proton state, the light-front eigenstate $|\psi\rangle$ is expanded as a superposition of Fock states,
\begin{equation}
|\psi_{\mathrm{proton}}\rangle
= \Phi^{qqq}\,|qqq\rangle
+ \Phi^{qqqg}\,|qqqg\rangle
+ \cdots ,
\end{equation}
where $\Phi^{qqq}$ and $\Phi^{qqqg}$ are the LFWFs corresponding to the $|uud\rangle$ and $|uudg\rangle$ Fock sectors, respectively, and the ellipsis ($\cdots$) denotes higher Fock components of the proton. In this work, the Fock space is truncated to these two lowest sectors.

The LF Hamiltonian adopted in this work is $P^- = P^-_{\rm QCD} + P^-_{\rm C}$, where $P^-_{\rm QCD}$ contains the relevant QCD interaction terms, and $P^-_{\rm C}$ is the confinement term that models the non-perturbative confinement effect \cite{Xu:2023nqv}. In the LF gauge $A^+ = 0$, the LF QCD Hamiltonian including one dynamical gluon is given by \cite{Xu:2023nqv}
\begin{align}
    P^-_{\mathrm{QCD}} =& \int d^2x^\perp dx^- \bigg\{\frac{1}{2} \bar{\psi} \gamma^+ \frac{m_0^2 + (i\partial^+)^2}{i\partial^+} \psi \notag \\
    &+\frac{1}{2} A_a^i \left[ m_g^2 + (i\partial^+)^2 \right] A_a^i 
    + g_s \bar{\psi} \gamma_\mu T^a A_a^\mu \psi \notag \\
    &+ \frac{1}{2} g_s^2 \bar{\psi} \gamma^+ T^a \psi \frac{1}{(i\partial^+)^2} \bar{\psi} \gamma^+ T^a \psi \bigg\},
\end{align}
where: $\psi$ is the quark field operator, $m_0$ is the bare quark mass, $m_g$ is the bare gluon mass, and $g_s$ is the strong coupling constant. Specifically, the first/second term in the Hamiltonian accounts for the kinetic energy of quarks/gluons, and the last two terms represent the vertex and instantaneous interaction.

The confinement term in the $|qqq\rangle$ sector is implemented following Ref. \cite{Li:2015zda} as
\begin{equation}
    P^-_{\rm C} P^+ = \frac{\kappa^4}{2} \sum_{i\neq j} \left[ \vec{r}_{ij\perp}^{\,2} - \frac{\partial_{x_i}\!\left(x_i x_j \partial_{x_j}\right)}{(m_i + m_j)^2} \right],
\end{equation}
where $\vec{r}_{ij\perp}$ is the transverse separation, and $\kappa$ is the confinement strength parameter. For the $|qqqg\rangle$ sector, no explicit confinement term is introduced; instead, the essential confinement behavior is captured by the restricted transverse basis and the inclusion of a massive gluon.

Within the BLFQ framework \cite{Vary:2009gt}, the proton state is expanded in a basis consisting of quarks and a gluon that reside in: (i) longitudinal plane waves (defined in a box of length $2L$ with antiperiodic boundary conditions for quarks and periodic boundary conditions for gluons), (ii) two-dimensional harmonic oscillator (2D-HO) functions $\Phi_{nm}(\vec{p}_\perp;b)$ (with $\vec{p}_\perp$ being the transverse momentum, $b$ the 2D-HO scale parameter, and $n,m$ the 2D-HO quantum numbers) \cite{Zhao:2014xaa}, and (iii) light-cone helicity spinors. As a result of these choices, the single-parton states in the basis are labeled by $\alpha_i = \{k_i,n_i,m_i,\lambda_i\}$, where $k$ is the longitudinal momentum quantum number (half-integer for quarks, integer for gluons, excluding the zero mode), and $\lambda$ is the helicity. For multi-parton sectors with more than one color-singlet configuration (e.g., $|qqqg\rangle$), an additional color label is required.

The basis expansion is truncated by two parameters: $N_{\rm max}$ and $K$. $N_{\rm max}$ imposes the transverse truncation condition $\sum_i (2n_i + |m_i| + 1) \le N_{\rm max}$, and $K$ fixes the longitudinal resolution with $x_i = k_i/K$, where $K=\sum_i k_i$. The parameters $b, N_{\mathrm{max}}$ and $K$ determine the IR and UV scales of the calculation \cite{Zhao:2014xaa}. Diagonalizing the LF Hamiltonian yields the proton LFWFs with helicity $\Lambda$, which can be expressed as
\begin{equation}
    \Psi^{N,\Lambda}_{\{x_i,\vec{p}_{\perp i},\lambda_i\}} = \sum_{\{n_i,m_i\}} \psi^{N}(\{\alpha_i\}) \prod_{i=1}^{N} \Phi_{n_i m_i}(\vec{p}_{i\perp}, b),
\end{equation}
where $N$ is the number of partons in the Fock sector ($N=3/4$ for $|uud\rangle/|uudg\rangle$), and $\psi^{N}(\{\alpha_i\})$ are the eigenvector components corresponding to the $N$-parton Fock sectors.

The Hamiltonian parameters employed in this work are summarized in Table \ref{paratabel}, with the truncation parameters fixed as $\{N_{\rm{max}}, K\} = \{9, 16.5\}$. These parameters are calibrated to reproduce the proton mass and its key electromagnetic properties \cite{Xu:2023nqv}. At the model scale, the proton state has a probability of approximately 44\% in the $|qqq\rangle$ sector and 56\% in the $|qqqg\rangle$ sector. The resulting LFWFs correspond to a low-resolution scale of $\mu_0^2 \sim 0.24 \pm 0.01~\text{GeV}^2$ \cite{Xu:2023nqv} and have been successfully applied to describe a wide range of proton observables, including electromagnetic and gravitational FFs, charge and mechanical radii, PDFs, generalized parton distributions (GPDs), transverse momentum-dependent PDFs (TMDs), as well as spin and orbital angular momentum distributions \cite{Xu:2023nqv,Yu:2024mxo,Zhu:2024awq,Zhang:2025nll,Lin:2024ijo,Lin:2023ezw,zhang2024twist,Nair:2025sfr,Zhu:2026azt}.

\begin{table}[h!]
\caption{\label{paratabel}
Summary of the model parameters~\cite{Xu:2023nqv}. All parameters have units of GeV except for $g_s$.}
\begin{ruledtabular}
\begin{tabular}{cccccccc}
$m_u$ & $m_d$ & $m_g$ & $\kappa$ & $m_f$ & $g_s$ & $b$ & $b_{\text{inst}}$ \\
\colrule
0.31 & 0.25 & 0.50 & 0.54 & 1.80 & 2.40 & 0.70 & 3.00 \\
\end{tabular}
\end{ruledtabular}
\end{table}

\section{\label{sec:3} Transverse color Lorentz force and overlap representation}
We start with the general matrix element containing the $\bar{q}Gq$ correlator, defined as \cite{aslan2019transverse}
\begin{equation}
    W^{\mu,\nu \rho}_{\Lambda^\prime \Lambda} (p^\prime, p) = \langle p',\Lambda' | \bar{\psi}(0) \gamma^{\mu} i g G^{\nu \rho}(0) \psi(0) | p, \Lambda \rangle,
\end{equation}
where $\psi(x)$ is the quark field operator, $G^{\nu \rho}$ is the gluon field strength tensor, and $|p, \Lambda \rangle$ denotes the proton state with four-momentum $p$ and helicity $\Lambda$. This matrix element can be parameterized into eight FFs. To extract a physical quantity, we take $\{\mu,\nu, \rho\} = \{+,+,j\}$, leading to the reduced twist-3 matrix element \cite{aslan2019transverse,PhysRevD.88.114502}
\begin{equation}
    d^q_2 = - \frac{\langle p,\Lambda| \bar{\psi}(0) \gamma^+ g G^{+y}(0) \psi(0) |p,\Lambda \rangle}{2M(p^+)^2 S^x},
\end{equation}
where $M$ is the proton mass and $S^x$ is the $x$-component of the proton polarization vector. This implies considering a proton transversely polarized along the $x$-direction, which is a superposition of opposite helicity states \cite{burkardt2002impact}
\begin{equation}
    |p_x \rangle = \frac{1}{\sqrt{2}} \Big( |p,\uparrow \rangle + |p,\downarrow \rangle \Big).
\end{equation}

The quantity $d^q_2$ is of particular interest because it does not only measure the response of the color electric and magnetic field to the polarization of the nucleon through the gluon field strength tensor \cite{Filippone:2001ux,PhysRevD.88.114502}
\begin{equation}
    G^{+y} \sim (E^y - B^x) = (\vec{E}+\vec{v} \times \vec{B})^y,
\end{equation}
but also rigorously related to the nucleon's second spin-dependent structure function $g_2$ via \cite{aslan2019transverse}
\begin{equation}
    \int_{-1}^1 dx x^2 \bar{g}_2^q (x) = \frac{d^q_2}{3},
\end{equation}
with $g_2^q(x)$ defined as \cite{wandzura1977sum}
\begin{align} \label{g2def}
    &g^q_2(x) = g^q_T(x) - g^q_1(x) = g^{q,WW}_2(x) + \bar{g}^q_2(x), \notag \\
    &g^{q,WW}_2(x) = -g^q_1(x) + \int_x^1 \frac{dy}{y}g^q_1(y),
\end{align}
where $g_1(x)$ is the twist-2 helicity PDF and $g_T(x)$ is the twist-3 spin-dependent PDF.

Considering the above constraints, the matrix element $W^{+,+j}_{\Lambda^\prime \Lambda}(p',p)$ reduces to five FFs. The parameterization reads \cite{aslan2019transverse}
\begin{widetext}
    \begin{align}
        W^{+, + j}_{\Lambda^\prime \Lambda} (p^\prime, p) &= \bar{u}(p',\Lambda') \left\{\frac{1}{M^2} [P^+ \Delta_\perp^j - P^\perp \Delta^+] \gamma^+ \Phi_1(t) + \frac{P^+}{M} i\sigma^{+j} \Phi_2(t) \right. \notag \\
        &+ \left. \frac{1}{M^3} i\sigma^{+\Delta} [P^+\Delta_\perp^j \Phi_3(t) - P^\perp\Delta^+ \Phi_4(t)] + \frac{P^+\Delta^+}{M^3} i\sigma^{j\Delta} \Phi_5(t)\right\} u(p,\Lambda),
    \end{align}
\end{widetext}
where $P^\mu = (p^\mu + p^{\prime \mu})/2$ is the average four-momentum of the proton, and $-t = \Delta^2 = (p^\prime - p)^2$ is the momentum transfer between the initial and final proton states. For a impact parameter space interpretation \cite{burkardt2002impact}, we take the limit $\xi \equiv -\Delta^+ / (2P^+) = 0$ (i.e., $\Delta^+ = 0$), in which only three terms in the above equation survive.

The transverse color Lorentz force is defined as
\begin{equation} \label{ftdef}
    F^j_{\Lambda^\prime \Lambda} (t) = \frac{i}{\sqrt{2} P^+} W^{+, + j}_{\Lambda' \Lambda} (p', p),
\end{equation}
and its transverse-plane distribution is obtained via the Fourier transform of Eq. (\ref{ftdef})
\begin{equation}
    \mathscr{F}^{j}_{\Lambda^\prime \Lambda} (\bm{b}_\perp) = \int \frac{d^2\Delta_\perp}{(2\pi)^2} e^{-i \bm{b}_\perp \cdot \bm{\Delta}_\perp} F^j_{\Lambda^\prime \Lambda} (\bm{\Delta_\perp}).
\end{equation}

We present the overlap representations of all relevant FFs below. For brevity, we adopt the following notations
\begin{align}
    [dx]_n &= \prod_{i=1}^n \frac{d x_{i} d^2 \vec{k}_{i}}{(16 \pi^3)^n} 16\pi^3 \delta \left(1-\sum x_i \right) \notag \\ 
    &\times \delta^2 \left(\sum \vec{k}_i \right) \delta (x-x_1),
\end{align}
where the subscript $1$ labels the struck quark.  We will now employ $\psi^{\Lambda}_n$ to denote the LFWF $\psi^{\Lambda}_{i=1,\cdots,n}(x_i,p_i,\lambda_i)$ with proton helicity $\Lambda$ and parton helicities $\lambda_i$; $[\Psi^{\Lambda^\prime \Lambda}_j] = \varepsilon_j \psi^{\Lambda^\prime \star}_{3} \psi^{\Lambda}_{4} - \varepsilon^\star_j \psi^{\Lambda^\prime \star}_{4} \psi^{\Lambda}_{3}$; and $[\gamma^+] = \bar{u} (p',\Lambda') \gamma^+ u (p,\Lambda)$ to encapsulate the helicity combinations of the struck quark, with $\varepsilon_\mu$ representing the gluon polarization vector. Here, the constraint $\delta^{\lambda'_2}_{\lambda_2} \delta^{\lambda'_3}_{\lambda_3} \delta^{\lambda'_4}_{\lambda_4}$ for spectator partons will be implicitly assumed, and $\Delta = \Delta_{1} + i\Delta_{2}$ will refer to the two-dimensional complex representation.

The momentum-space overlap representations of the FFs can then be written
\begin{align}
    \Phi_1^j(t) &= \int [dx]_4 \frac{M^2 [\gamma^+]}{\sqrt{2}(P^+)^2 \Delta_j} [\Psi^{\uparrow \uparrow}_j], \label{phi1} \\
    \Phi_2^j(t) &= \int [dx]_4 \frac{\sqrt{2} i M [\gamma^+]}{4 (P^+)^2 \epsilon^{kj}_\perp \Delta_k} \Big(\Delta[\Psi^{\uparrow \downarrow}_j] + \Delta^\star[\Psi^{\downarrow \uparrow}_j] \Big), \label{phi2} \\
    \Phi_3^j(t) &= \int [dx]_4 \frac{\sqrt{2} i^j M^3 [\gamma^+]}{8 (P^+)^2 \Delta_1 \Delta_2} \Big((-)^{j+1} [\Psi^{\uparrow \downarrow}_j] + [\Psi^{\downarrow \uparrow}_j] \Big), \label{phi3}
\end{align}
where $\epsilon_\perp^{kj}$ is the transverse Levi-Civita symbol, and $j$ takes values 1 or 2 (the choice of $j$ does not affect $\Phi_k(t)$). The repeated index $j$ is not summed over. The impact parameter space components of the transverse color Lorentz force distribution are derived as
\begin{align}
    \mathscr{F}^j_{\Lambda^\prime \Lambda,1} (\mathbf{b}_\perp) =& \frac{i}{\sqrt{2}P^+} \int \frac{d^2\Delta_\perp}{(2\pi)^2} e^{-i \bm{b}_\perp \cdot \bm{\Delta}_\perp} \notag \\
    &\times \bar{u}(p^\prime,\Lambda^\prime) \left[\frac{P^+\Delta_\perp^j}{M^2}\gamma^+ \Phi_1(t) \right] u(p,\Lambda) \notag \\
    =& -\frac{P^+}{M^2} \frac{\partial}{\partial b_j} \tilde{\Phi}_1 (b^2), \\
    \mathscr{F}^j_{\Lambda^\prime \Lambda,2} (\mathbf{b}_\perp) =& \frac{i}{\sqrt{2}P^+} \int \frac{d^2\Delta_\perp}{(2\pi)^2} e^{-i \bm{b}_\perp \cdot \bm{\Delta}_\perp} \notag \\
    &\times \bar{u}(p^\prime,\Lambda^\prime) \left[\frac{P^+}{M} i\sigma^{+j} \Phi_2(t) \right] u(p,\Lambda) \notag \\
    =& \frac{2 P^+}{M} \tilde{\Phi}_2 (b^2) \delta^j_2, \\
    \mathscr{F}^j_{\Lambda^\prime \Lambda,3} (\mathbf{b}_\perp) =& \frac{i}{\sqrt{2}P^+} \int \frac{d^2\Delta_\perp}{(2\pi)^2} e^{-i \bm{b}_\perp \cdot \bm{\Delta}_\perp} \notag \\
    &\times \bar{u}(p^\prime,\Lambda^\prime) \left[\frac{P^+\Delta_\perp^j}{M^3} i\sigma^{+\Delta} \Phi_3(t) \right] u(p,\Lambda) \notag \\
    =& -\frac{4 P^+}{M^3} \left( \frac{\partial^2}{\partial b_1 \partial b_2} \delta^j_1 + \frac{\partial^2}{\partial b_2^2} \delta^j_2 \right) \tilde{\Phi}_3 (b^2),
\end{align}
where $\tilde{\Phi}_i$ are the Fourier transforms of the corresponding momentum-space FFs.

Notably, $\Phi_4$ and $\Phi_5$ do not contribute due to $\Delta^+=0$. From Eq.~(\ref{phi1}), $\Phi_1$ contributes only for identical initial and final nucleon helicities, implying that the first force component, $\mathscr{F}_1$, is insensitive to the nucleon polarization. In contrast, Eqs.~(\ref{phi2})--(\ref{phi3}) show that $\Phi_2$ and $\Phi_3$ require a nucleon helicity flip, and consequently $\mathscr{F}_2$ and $\mathscr{F}_3$ are polarization dependent. It has been shown that $\mathscr{F}_2$ corresponds to the spatial distribution of the Sivers force~\cite{sivers1991hard}.

\section{\label{sec:4} Numerical results}
\begin{figure*}[htbp]
    \centering
    \subfloat[ ]{\includegraphics[scale=0.65]{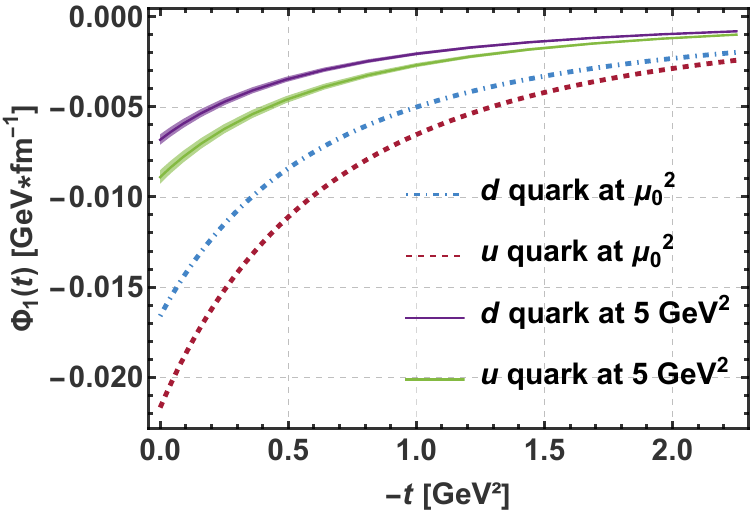}} \hspace{.3in}
    \subfloat[ ]{\includegraphics[scale=0.65]{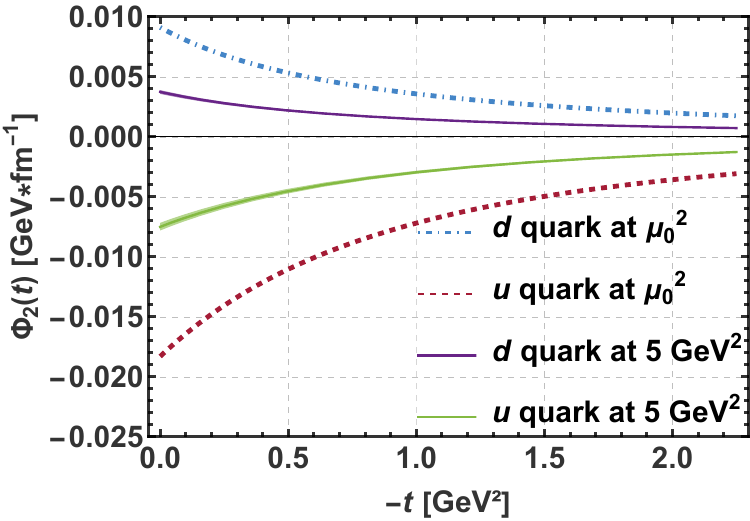}} \vspace{-.3in} \\
    \subfloat[ ]{\includegraphics[scale=0.65]{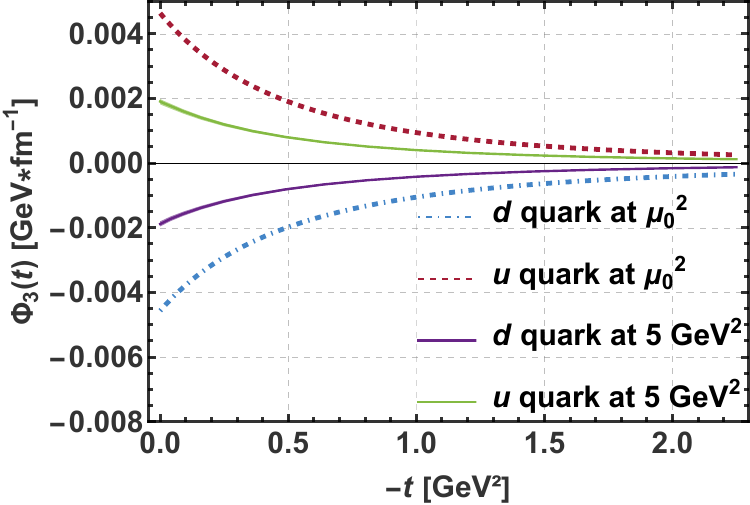}}
    \captionsetup{justification=raggedright}
    \caption{Plots of the FFs $\Phi_1, \Phi_2$ and $\Phi_3$ as a function of $-t$ for quarks inside a proton. The blue dash-dotted line stands for $d$ quark at the initial scale and purple solid line stands at 5 GeV$^2$. The red dashed line represents results for $u$ quark at the initial scale and the green solid line at 5 GeV$^2$.}
    \label{momspace}
\end{figure*}

In Fig.~\ref{momspace}, we present the FFs $\Phi_1$, $\Phi_2$, and $\Phi_3$ for $u$ and $d$ quarks in the proton as functions of the momentum transfer $-t$, ranging from $0$ to $2.25~\mathrm{GeV}^2$. To facilitate comparison at a common scale, we use the \texttt{snowflake} package~\cite{rodini2024numerical} to evolve our results to $5~\mathrm{GeV}^2$. The associated uncertainty bands arise from the variation of the initial scale, $\mu_0^2 = 0.24 \pm 0.01~\mathrm{GeV}^2$.


For $\Phi_1$ [panel (a)], both quark flavors exhibit negative distributions, characterized by a rapid increase in magnitude at small $-t$, followed by a gradual saturation at larger $-t$. The magnitudes of the $u$- and $d$-quark contributions are similar. In contrast, $\Phi_2$ [panel (b)] and $\Phi_3$ [panel (c)] display markedly different behaviors: the $u$- and $d$-quark contributions have opposite signs, not only between the two quark flavors but also between the two FFs. The magnitude of $\Phi_1$ for the $u$ quark is slightly larger than that for the $d$ quark, whereas in $\Phi_2$ the $u$-quark contribution is approximately twice as large as that of the $d$ quark, and in $\Phi_3$ the two magnitudes are comparable. The evolved results preserve the qualitative features of the unevolved distributions.


Taking the forward limit $\Delta \to 0$, one obtains the closely related twist--3 quantity $d_2^q$,
\begin{equation}
d_2^q = -\frac{F(0)}{M^2} = -\frac{\Phi_2(0)}{M^2}.
\end{equation}
Using the light-front wave functions obtained within our BLFQ framework, we find
$d_2^d = -0.00996$ ($d_2^u = 0.0200$) at the initial scale, and
$d_2^d = -0.00409$ ($d_2^u = 0.00820$) at $5~\mathrm{GeV}^2$.
The corresponding nucleon results are
$d_2^p = 0.00778$ ($d_2^n = -0.00221$) at the initial scale, and
$d_2^p = 0.00319$ ($d_2^n = -0.000906$) at $5~\mathrm{GeV}^2$.
Similar values have been obtained in various other model calculations, as summarized in Table~\ref{d2tabel}.
Using the extracted $d_2$ values and $M^2 = 4.855~\mathrm{GeV/fm}$, the magnitude of the color--Lorentz force is estimated to be
$F_u \approx -0.04~\mathrm{GeV/fm}$ and
$F_d \approx 0.02~\mathrm{GeV/fm}$ at $5~\mathrm{GeV}^2$.

\begin{table*}
\caption{\label{d2tabel}
The results contain $d_2^p$ (for proton) and $d_2^n$ (for neutron) compared here are at 5 GeV$^2$ except for ILM and QCDSF which are evaluated at 4 GeV$^2$. The Hall A \cite{flay2016measurements} gives -0.00035(83)(69)(7) for neutron at 4.3 GeV$^2$.}
\begin{ruledtabular}
\begin{tabular}{ccccccccc}
        & BLFQ & RQCD\cite{burger2022lattice} & LQCD\cite{gockeler2005investigation} & HERMES\cite{hermes2012measurement} & c.m. bag\cite{song1996polarized} & Global fit\cite{Vladimirov:2025qrh} & ILM\cite{liu2025color} & QCDSF\cite{crawford2025transverse} \\
\colrule
$Q^2$(GeV$^2$) & 5.0 & 5.0 & 5.0 & 5.0 & 5.0 & 5.0 & 4.0 & 4.0 \\
\colrule
$d_2^p$ & 0.00319 & 0.0105(68) & 0.004(5) & 0.0148(96)(48) & 0.0174 & 0.00347(137) & 0.00550 & 0.046(7)(16) \\
\colrule
$d_2^n$ & -0.000906 & -0.0009(70) & -0.001(3) & - & -0.00253 & -0.00217(63) & 0.000198 & 0.023(5)(8)
\end{tabular}
\end{ruledtabular}
\end{table*}

\begin{figure*}[htbp]
    \centering
    \subfloat[ ]{\includegraphics[scale=0.28]{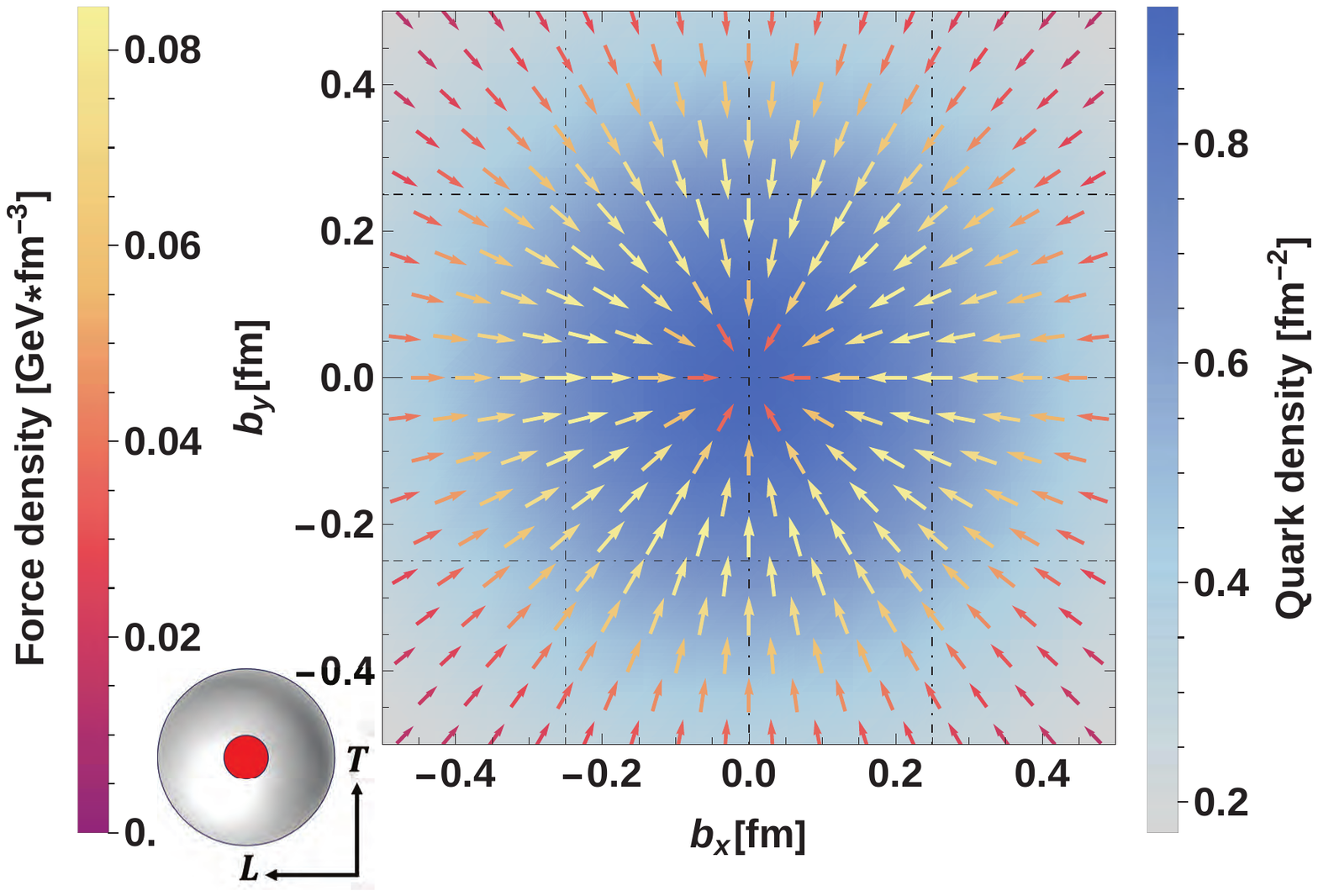}} \hspace{.3in}
    \subfloat[ ]{\includegraphics[scale=0.28]{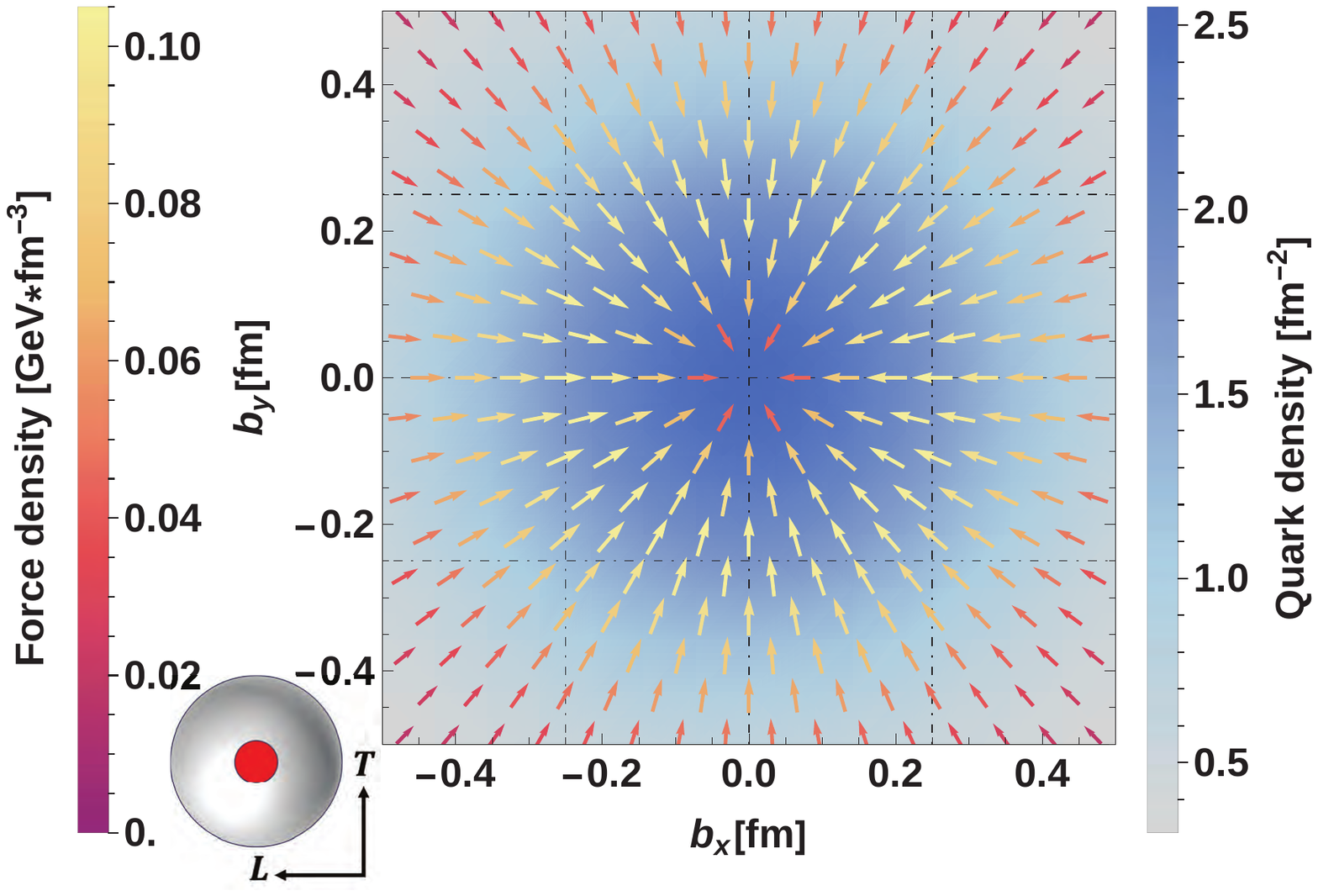}} \\
    \subfloat[ ]{\includegraphics[scale=0.28]{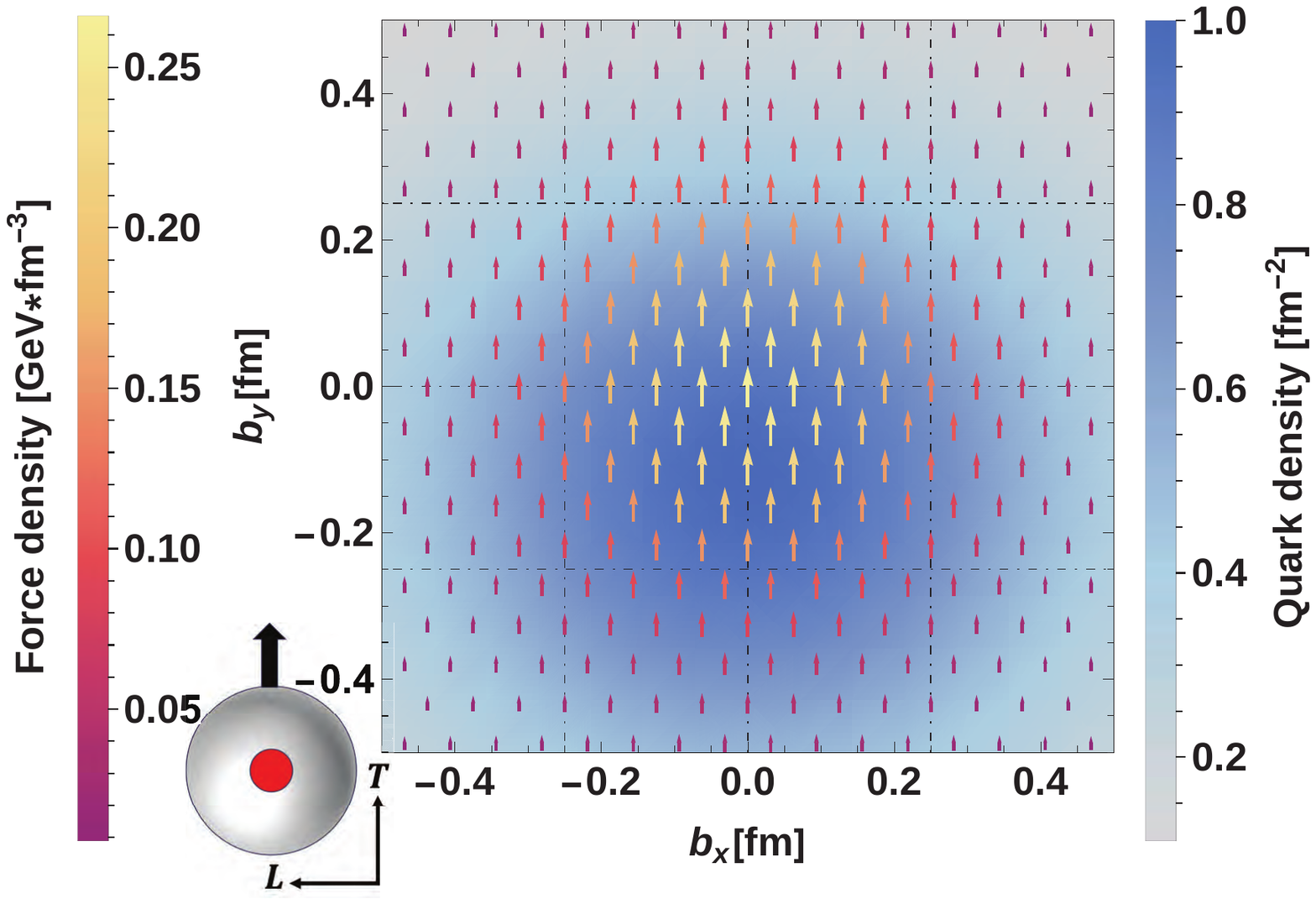}} \hspace{.3in}
    \subfloat[ ]{\includegraphics[scale=0.28]{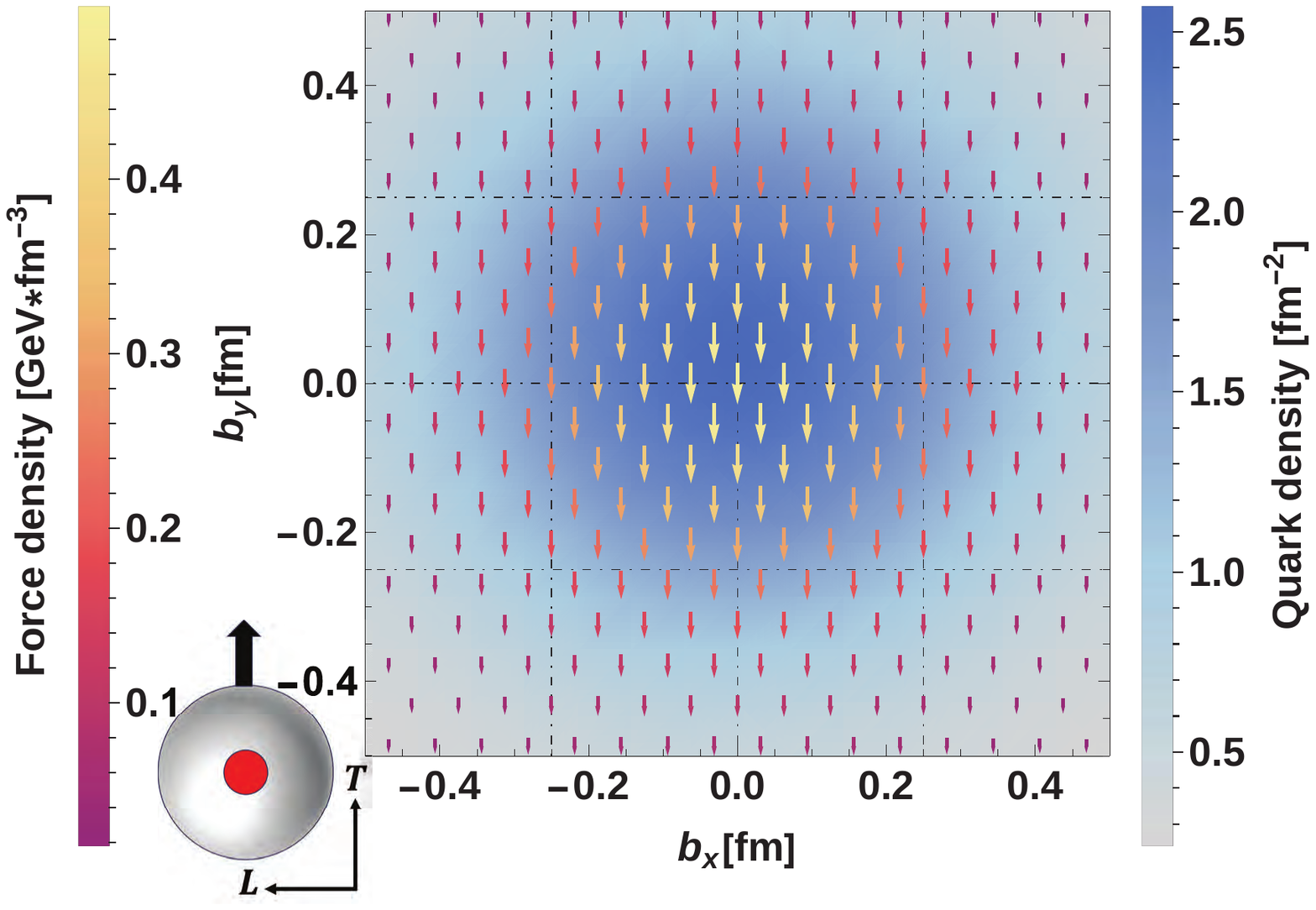}} \\
    \subfloat[ ]{\includegraphics[scale=0.28]{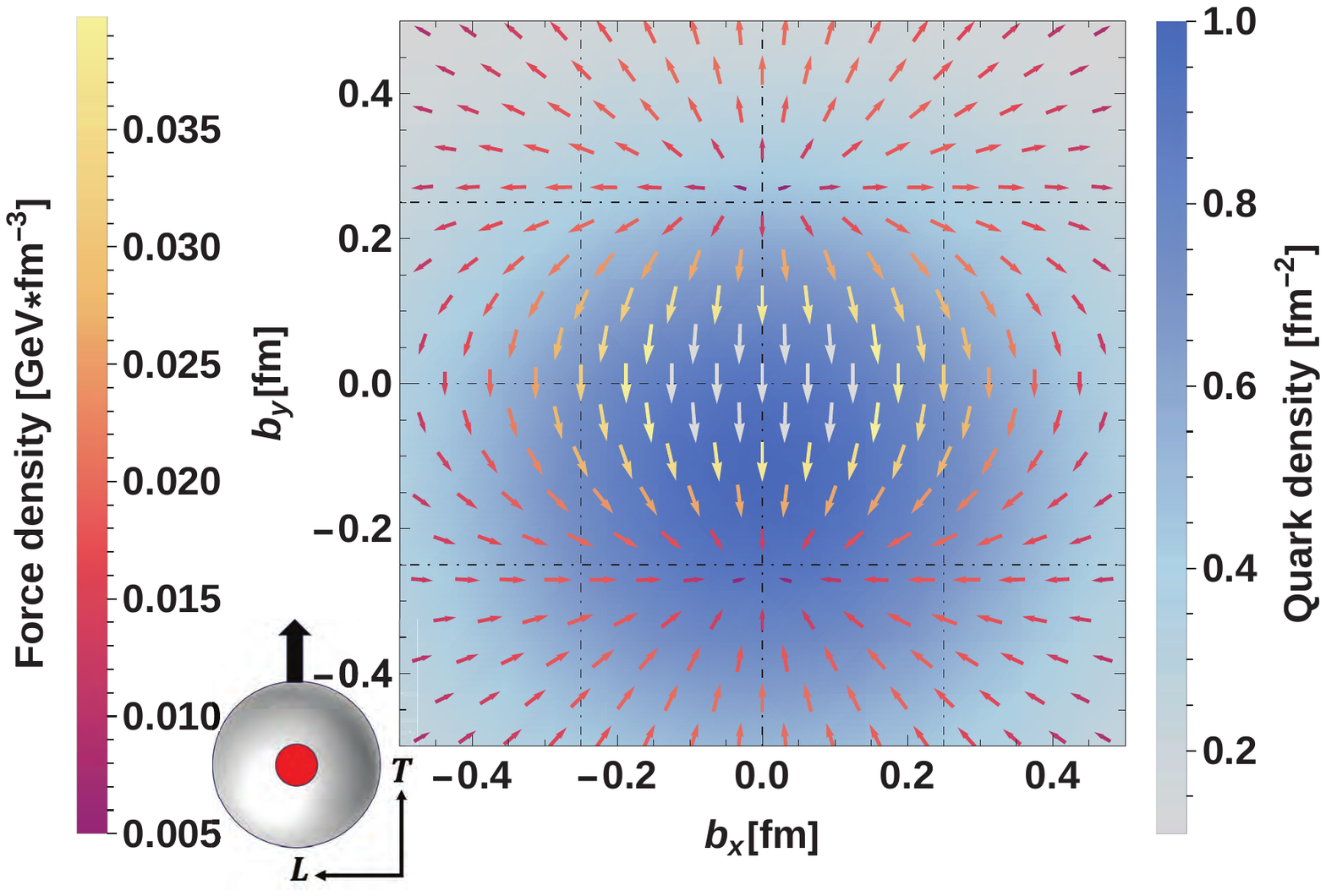}} \hspace{.3in}
    \subfloat[ ]{\includegraphics[scale=0.28]{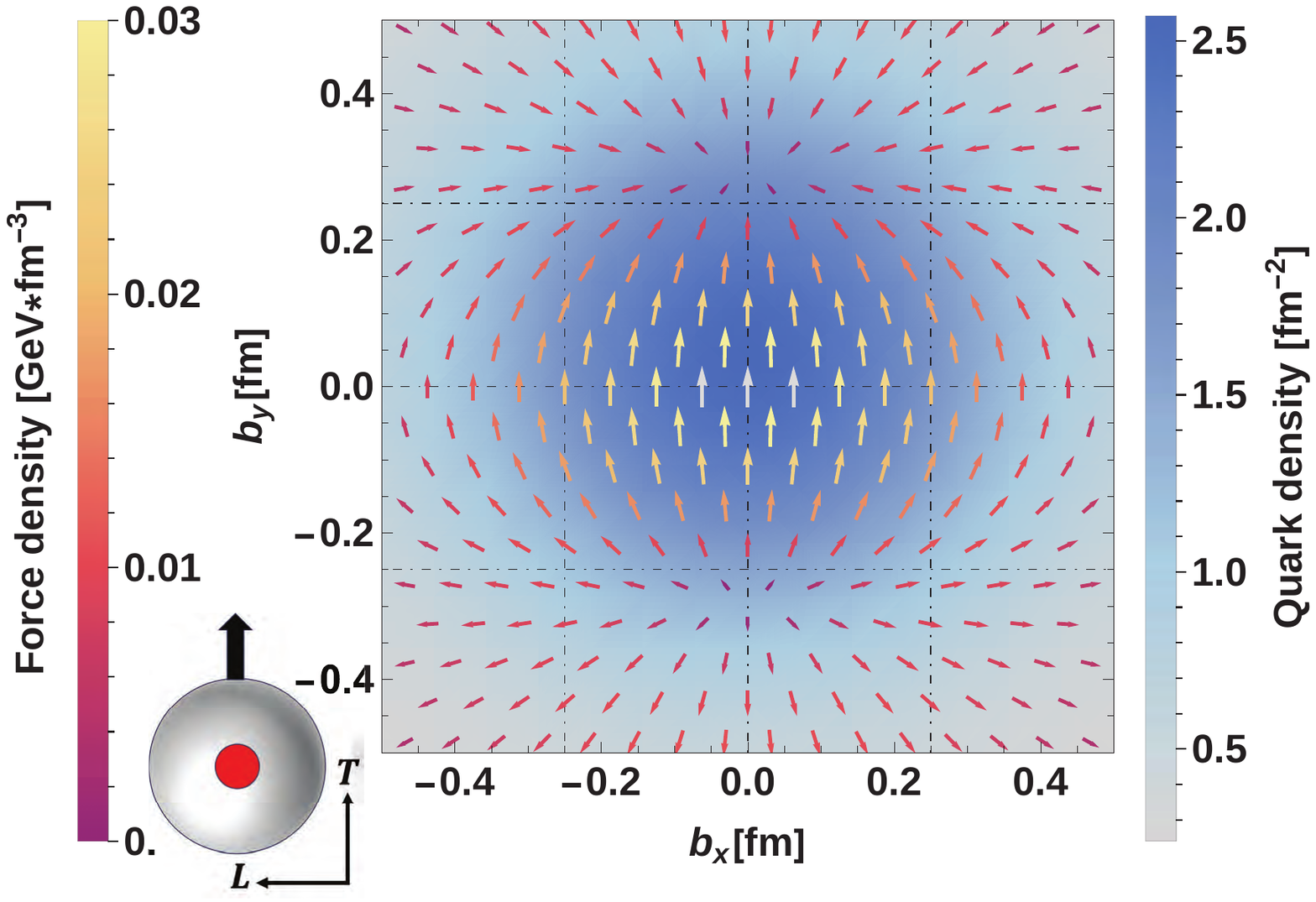}}
    \captionsetup{justification=raggedright}
    \caption{Plots of the force components $\mathcal{F}_1, \mathcal{F}_2$ and $\mathcal{F}_3$ distributions in coordinate space for quarks together with quark density distributions. Left panel for down quark and right panel for up quark. From top to down they are $\mathcal{F}_1, \mathcal{F}_2$ and $\mathcal{F}_3$, respectively. The gray sphere in the lower-left corner denotes the proton, the black arrow represents the proton polarization direction, where L stands for longitudinal polarization and T for transverse polarization, and the red dot stands for unpolarized quark.}
    \label{plotforce}
\end{figure*}

In Fig. \ref{plotforce}, we present the results of the force component distributions in impact parameter space, overlaid on the quark density plots. From top to bottom, the panels correspond to $\mathscr{F}_1, \mathscr{F}_2$, and $\mathscr{F}_3$, with $d$-quark distributions on the left and $u$-quark distributions on the right. For the first component ($\mathscr{F}_1$), both $u$ and $d$ quarks exhibit radially inward attractive forces toward the center. Notably, this is not a smooth increase; instead, the force reaches a maximum value at around $|\bm{b}_\perp|\sim 0.25$ fm and tends to 0 near the central ($|\bm{b}_\perp|=0$).
For the second component ($\mathscr{F}_2$), both $u$ and $d$ quarks exhibit pronounced alignment along the $b_y$ axis, with their maximum magnitudes concentrated in the central region and opposite signs between the two quark flavors. For the third component ($\mathscr{F}_3$), a dipole structure is observed with maximum values gathered at the center: for $d$-quark, there is a repulsive pole at $(b_x,b_y) \sim (0, 0.27)$ and an attractive pole at $(b_x,b_y) \sim (0, -0.27)$, leading to a force direction from the upper region to the lower region; the $u$-quark exhibits the opposite sign.

The two-dimensional Fourier transforms of these FFs $\Phi_1$, $\Phi_2$, and $\Phi_3$ reveal strong, localized forces acting on the struck quark. For an unpolarized proton, the force distribution exhibits a central restoring force, whereas for a transversely polarized proton it shows a pronounced downward force in the region of highest quark density. From a phenomenological perspective, these results offer a complementary interpretation of the Sivers asymmetry observed in transversely polarized SIDIS experiments. Overall, the two-dimensional force distributions provide an intuitive and transparent visualization of the underlying mechanisms responsible for single-spin asymmetries.

The background in these plots represents the quark density, in units of fm$^{-2}$, obtained by integrating over the longitudinal momentum fraction of the Fourier transforms of the twist-2 GPDs $H(x,\bm{b})$ and $E(x,\bm{b})$, or equivalently of the Dirac and Pauli form factors~\cite{Burkardt:2002hr,Pasquini:2007xz,Maji:2017ill,Mondal:2015uha}. These densities correspond to a proton polarized along the $x$ direction. By normalizing the force density with respect to the quark density, one obtains the average local force strength in a given region. The force distributions shown in Fig.~\ref{plotforce} are in good agreement with the results reported in Ref.~\cite{aslan2019transverse}.

\section{\label{sec:5} Summary and conclusions}
This study systematically investigates the transverse color Lorentz force and its intrinsic connection to the nucleon twist-3 spin–dependent structure function $g_2$ within the BLFQ framework. First, we compute the momentum-space force form factors $\Phi_1$, $\Phi_2$, and $\Phi_3$ for $u$ and $d$ quarks in a transversely polarized proton. The results reveal clear flavor-dependent features: $\Phi_1$ is negative for both quark flavors, with its magnitude saturating at large $-t$, while $\Phi_2$ and $\Phi_3$ exhibit opposite sign patterns for $u$ and $d$ quarks. Moreover, the magnitudes of $\Phi_1$, $\Phi_2$, and $\Phi_3$ for $u$ quarks are approximately $1.5$, $2$, and $1$ times larger, respectively, than those for $d$ quarks.

Second, we extract the twist-3 matrix element $d_2^q$ from the forward limit and obtain numerical estimates for both quark flavors. Since $d_2$ is predicted to be intrinsically small, its experimental determination will likely be subject to sizable uncertainties. Nevertheless, our theoretical predictions are consistent with existing lattice QCD results and phenomenological extractions.

Third, the force distributions in impact-parameter space exhibit distinct spatial structures for each component. The component $\mathscr{F}_1$ corresponds to a radially attractive force that increases from the periphery toward intermediate distances and then decreases to zero near the center. The component $\mathscr{F}_2$ is oriented along the $b_x$ direction, with its sign depending on the quark flavor, while $\mathscr{F}_3$ displays a dipole-like pattern characterized by repulsive and attractive poles and also shows flavor dependence. Accordingly, $\mathscr{F}_2$ and $\mathscr{F}_3$ require a proton helicity flip, whereas $\mathscr{F}_1$ is insensitive to nucleon polarization.

Future work will focus on expanding the number of Fock-sectors and increasing the basis truncation parameters ($N_{\rm max}$ and $K$) to achieve higher numerical precision. We also plan to extend this framework to other hadron and nuclear systems and to investigate the color Lorentz force acting on transversely polarized quarks~\cite{PhysRevD.88.114502}. It should be noted, however, that experimental access to this latter observable is expected to be more challenging compared to twist-3 spin–dependent structure functions.

\begin{acknowledgments}
We thank Jiangshan Lan, Zhimin Zhu and Jiatong Wu for many helpful discussions. 
This work is supported by the National Natural Science Foundation of China under Grant No. 12305095, No.12375143, and No. 12250410251, by the Gansu International Collaboration and Talents Recruitment Base of Particle Physics (2023-2027), by the Senior Scientist Program funded by Gansu Province, Grant No. 25RCKA008.
C. M. is supported by new faculty start up funding the Institute of Modern Physics, Chinese Academy of Sciences, Grants No. E129952YR0. 
X. Z. is supported by Key Research Program of Frontier Sciences, Chinese Academy of Sciences, Grant No. ZDBS-LY-7020, by the Natural Science Foundation of Gansu Province, China, Grant No. 20JR10RA067, by the Foundation for Key Talents of Gansu Province, by the Central Funds Guiding the Local Science and Technology Development of Gansu Province, Grant No. 22ZY1QA006, by international partnership program of the Chinese Academy of Sciences, Grant No. 016GJHZ2022103FN, by the Strategic Priority Research Program of the Chinese Academy of Sciences, Grant No. XDB34000000.
J. P. V. is supported by the Department of Energy under Grant No. DE-SC0023692.  A portion of the computational resources were also provided by Taiyuan Advanced Computing Center.
\end{acknowledgments}

\bibliographystyle{unsrt}
\bibliography{main}

\end{document}